# The use of a simple digital weather station (not only) in teaching physics

**M Hruška[1*] and M Plesch[1]**

[1] Department of Physics, Faculty of Natural Sciences, Matej Bel University in Banská Bystrica, Slovakia

*E-mail: martin.hruska@umb.sk

**Abstract.** One of key goals of contemporary physics (and, realistically, STEM) education is to develop students' science literacy and critical thinking skills. In this paper, we present the construction and use of several versions of a simple school-based digital weather station that students can use to measure fundamental physical quantities (temperature, pressure, air humidity, light intensity) as part of school activities. The weather stations were constructed at our workplace using an Arduino microcontroller, BBC micro: bit, and the school measurement system Coach. This paper proposes not only the design and related programming of the weather stations but also how students can collect, analyse, and interpret measured data, thereby learning scientific methods and developing science literacy and critical thinking. This hands-on approach also develops students' experimental skills, emphasises the cross-curricular relationships between physics, computer science and geography, and teaches them to work with accurate data in the context of real environmental problems.

## 1. Introduction

As is well known, the development of critical thinking and scientific literacy are among the priority goals of contemporary STEM education. According to the OECD definition (e.g., within the PISA program), scientific literacy primarily refers to the ability to engage with science-related issues and with the ideas of science as a reflective citizen [1]. It encompasses not only knowledge of scientific disciplines but also the ability to analyze evidence, interpret data, evaluate arguments, and make informed decisions based on evidence.

Currently, the PISA 2025 study is transitioning to a broader concept known as the science framework. In developing the 2025 framework, two previous competencies ("Evaluate and design scientific enquiry" and "Interpret data and evidence scientifically") were merged into one: 'Construct and evaluate designs for scientific enquiry and interpret scientific data and evidence critically' [2]. Suppose we focus on the current goals of science education, as defined in the State Education Program, specifically for the educational area of Man and Nature (which includes physics along with chemistry and biology) [3]. In that case, the primary objective is to equip students with the fundamentals of science literacy, enabling them to formulate scientifically informed judgments and apply the knowledge they have acquired to solve problems effectively. In order to achieve this goal, students must develop cognitive skills that are also among the basic indicators for assessing critical thinking: the ability to formulate hypotheses, obtain relevant information and data, analyze and evaluate them, and then formulate conclusions based on arguments; and self-regulation, or the ability to consciously monitor one's cognitive activities and the results obtained in order to recognize or correct them [4].

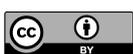

Based on the above, without critical thinking, science literacy would merely be a passive memorization of facts. On the contrary, critical thinking gives science literacy a deeper dimension – it allows students to think like scientists, systematically investigate phenomena, and test hypotheses.

Therefore, from the perspective of teaching physics in schools, and in accordance with the State Educational Program, it is essential to focus on using various forms of active learning and investigations so that students have as many opportunities as possible to learn selected (most often experimental) methods of investigating physical phenomena through hands-on activities. In teaching, the greatest attention is thus paid to the independent work of students – activities that focus on constructing new knowledge while also providing students with the opportunity to learn about the relationship between the development of natural sciences and the advancement of technology and its applications.

Experience suggests that such a holistic approach to STEM education has both advantages and disadvantages. For example, a 2024 study conducted in Norwegian secondary schools showed that teachers perceived such long-term STEM activities as very beneficial. The primary reason for their limited use is the traditional education system, which is based on adherence to the curriculum and does not provide sufficient time for teaching or teacher training, presenting several logistical obstacles [5]. We encountered similar problems in Slovak secondary schools when implementing inquiry-based project activities with interdisciplinary overlap (STEM) in the previous period [6].

The active approach to using digital technologies by current and upcoming generations of students can be applied not only in physics teaching but also in other subjects within STEM education. Therefore, Arduino or BBC micro:bit microcontrollers are often used in education, and in references, we can find several examples of designs for various measuring and sensing devices, including devices for measuring atmospheric physical properties, as well as ideas for their use in physics teaching [7-11]. On the other hand, for the simpler BBC micro:bit microcontroller, there are several educational kits that, once connected, can function as simple weather stations (e.g., SparkFun micro:climate kit for micro:bit or Smart Climate Kit (Elecfreaks)).

Based on the above, we were inspired by these and similar projects. However, instead of adopting an existing online design, we decided to develop our own versions of digital weather stations explicitly focused on educational purposes. The main reasons were simplicity, affordability, and didactic flexibility. Unlike many online projects that aim to measure a wide range of meteorological quantities and often require complex hardware, our approach deliberately focuses only on a few key physical quantities – temperature, pressure, humidity, and light intensity. This focus ensures that the stations remain technically simple and inexpensive while still providing rich opportunities for students' inquiry. At the same time, we designed parallel versions of the weather station using different hardware platforms (Arduino, BBC micro:bit, and the school measurement system Coach with VinciLab data logger). It enables the achievement of comparable results across different technologies, depending on the school's equipment and educational goals.

## 2. Design of digital weather stations

As mentioned above, three types of weather stations with different platforms for measuring basic variables – temperature, air pressure, relative humidity, and light intensity – were designed and built at the authors' workplace. Two are based on the low-cost Arduino Uno and BBC micro:bit microcontrollers, in combination with sensors connected via an I$^2$C bus and a simple LCD. The third was created using the Coach system for STEM education and a VinciLab data logger to collect and analyse data in the Coach 7 software or MS Excel environment.

Each of these platforms enables students to measure selected physical quantities over a more extended period of several days and process the obtained data, for example, using the Microsoft Excel spreadsheet program. To keep the design and programming simple, we decided to store data from weather stations to a computer via a serial port using USB (in the case of Arduino Uno and BBC micro:bit weather stations) or to transfer it using a USB flash drive after the measurement is complete (VinciLab weather station).

The weather stations were designed to be used at various levels depending on the difficulty of education, time allocation, and educational goals. Weather stations based on Arduino Uno and BBC micro:bit microcontrollers enable students to learn about the basic principles of sensor and display operation, communication via the I$^2$C interface, and microcontroller programming. Assuming that our primary goal is only to teach students how to acquire, process and evaluate data and monitor the weather or the physical state of the atmosphere, it is appropriate to work with ready-made weather stations or use a VinciLab data logger and suitable sensors for data collection, since their installation is not technically or time-consuming.

In the next part of the article, we will describe in more detail the Arduino and BBC micro:bit versions of the prepared digital weather station, whose design enables the development of students' ability to algorithmize a given problem and enhance their programming skills. We will also briefly discuss the basic features of the weather station version, which utilises the VinciLab data logger.

*2.1 Arduino and BBC micro:bit weather stations*

Arduino and BBC micro:bit weather stations (Figure 1a, b) were designed to be constructed relatively easily using several affordable modules (LCD, three sensors) while also ensuring high reliability in long-term measurements and ease of operation and installation.

Based on our own experience, we found that the combination of a four-line LCD and three sensors yielded the best results. The BH170 sensor measures light intensity in lux, the BMP180 or BMP280 sensor measures barometric pressure, and the HTU21D sensor records relative humidity and air temperature. The display allows us to view meteorological data easily. However, we can also record it in a text file by connecting the weather station to a computer via a USB cable.

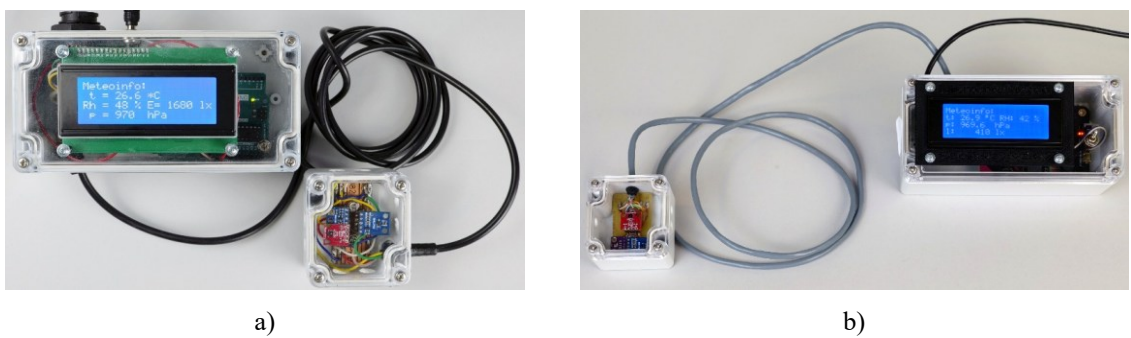

a) b)

**Figure 1.** Simple weather station with Arduino Uno (a) and BBC Micro:bit (b) microcontrollers.

*Arduino Uno/BBC micro:bit microcontrollers*

Both microcontrollers are suitable for first-time users who want to create their projects. There are many different projects available for them on publicly accessible forums (instructables.com, projecthub.arduino.cc, github.com, hackster.io, etc.).

*20 x 4 LCD*

It is a standard LCD with blue backlighting, 20 characters per line and four lines. The display also features an I$^2$C bus, enabling connection to microcontrollers via four wires: SLC, SDA, GND, and either 3.3 V or 5 V DC. The potentiometer on the I$^2$C interface module on the back of the display can be used to adjust its contrast depending on the power supply. At the same time, it is possible to switch the display backlight – in our case, we connected the contacts to a switch on the top cover of the box, allowing us to turn off the display backlight, especially at night.

*BH1750 light intensity sensor*

The light intensity sensor (Figure 2a) communicates via the I$^2$C interface, converting the measured light intensity into a digital output in the form of a numerical value. The sensor is adapted to the spectral characteristics of the human eye. As shown in Figure 3 below, we used a 3.3 V voltage supplied directly by the microcontroller to power this and the other two measuring modules.

*BMP180 and BMP280 barometric pressure sensors*

The BMP180 and BMP280 barometric pressure sensors (Figure 2b, c) from Bosch can measure both temperature and pressure. The typical pressure measurement accuracy is ±1 hPa, with the sensors measuring pressure in the range of 300 hPa to 1100 hPa. The sensors also communicate via the I$^2$C interface, so only four wires are needed to connect them to the microcontroller.

*Relative humidity and temperature sensor HTU21D*

The HTU21D sensor (Figure 2d) can measure temperatures ranging from -40 °C to +125 °C and relative humidity from 0% to 100%. The temperature accuracy is typically ±0.3 °C, with the most accurate measurements achieved in the range of 5 °C to 60 °C. For humidity, the accuracy is ±1%, and the sensor measures most accurately within the range of 10% to 90%. The manufacturer recommends a supply voltage for this module in the range of 3.3 V to 5 V.

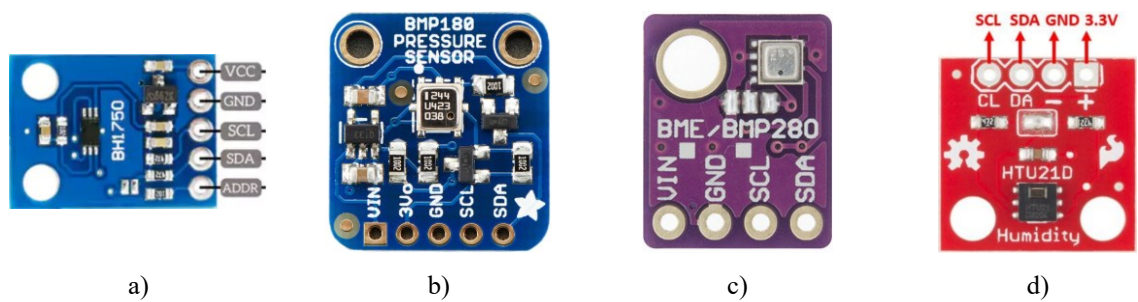

a) b) c) d)

**Figure 2a, b, c, d.** The light intensity sensor BH1750 (a), the pressure sensors BMP180 (b) and BMP280 (c) and the temperature and relative humidity sensor HTU21D (d).

*Description of the connection and design of the weather stations*

The LCD, as well as all three sensors mentioned above, use the I$^2$C serial interface to communicate with microcontrollers. The wire marked SCL (Serial Clock) is used to clock the communication, while the second wire marked SDA (Serial Data) is used to transfer data. Additionally, the GND and 3.3 V power wires must be connected to the devices. One device that controls communication must be of the Master type (typically a microcontroller), while the other devices are of the Slave type. Each device is identified by its unique address when communicating via I$^2$C [12]. The wiring diagram for the Arduino Uno and BBC micro:bit weather stations, featuring an LCD and sensors, is shown in Figure 3a, b.

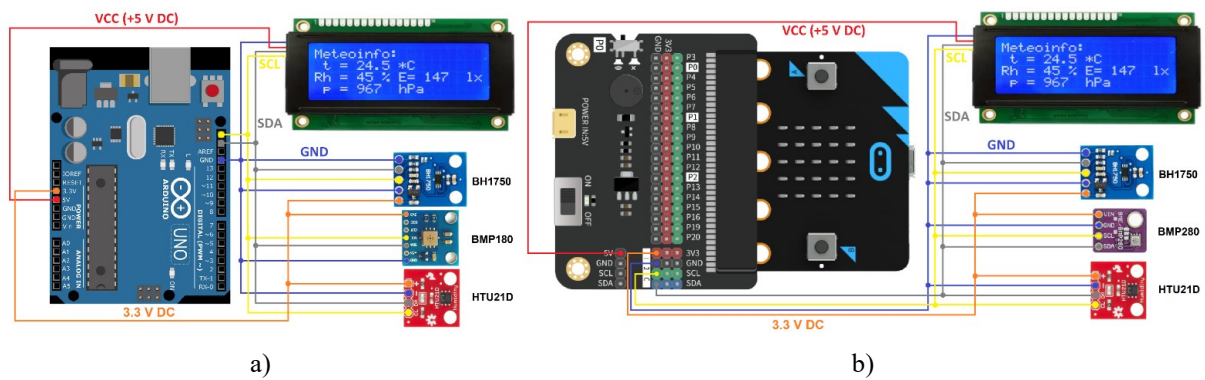

a)  b)

**Figure 3.** Wiring diagram of Arduino (a) and BBC Micro:bit (b) weather stations.

Due to the use of the I$^2$C bus and the same 3.3 V DC power supply, the sensors can be mounted on a suitable printed circuit board (the spacing between the sensor contact holes should be 0.1", which corresponds to 2.54 mm), interconnect them according to the diagram in Figure 3a, b, and fit them into a small mounting box with a transparent cover (Figure 4a, b), which can be placed outdoors, for example behind a window. The box with the sensors is connected to the Arduino using a four-wire cable with SCL, SDA, GND, and 3.3 V DC wires.

The weather stations themselves (Figure 1) were also placed in a mounting box with a transparent cover.

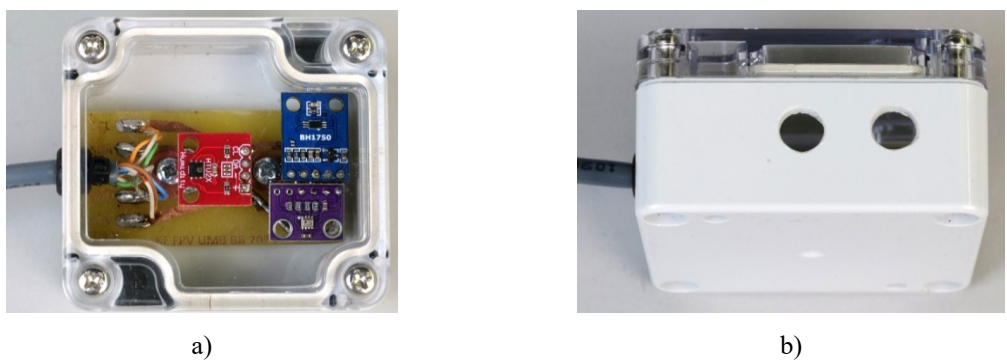

a)  b)

**Figure 4a, b.** Mounting sensors on a printed circuit board and installing the printed circuit board with sensors in a mounting box featuring a transparent top cover (a). View of the drilled holes from the side of the box (b).

*Programming a weather station with Arduino Uno*

In the Arduino IDE, the program first loads the necessary libraries and specifies the connected devices. In *setup()*, we initialise communication, sensors, and the LCD display, and write fixed labels. The *loop()* section then repeatedly reads sensor values, converts pressure to hPa, displays the results on the LCD, and sends them to the computer. Using *delay()*, the program updates and records pressure (hPa), temperature (°C), humidity (%), and light intensity (lx) about once per minute.

*Programming a weather station with BBC micro:bit*

For a weather station with a BBC micro:bit microcontroller, it is best to program online in the MicroPython environment via the website https://python.microbit.org/v/3. Since MicroPython is a higher-level language, programming in it is clearer and more understandable, which makes it easier to use in teaching. However, the principles of working with sensors, communicating via the I$^2$C bus, or processing and interpreting data remain the same as when programming in C/C++ for the Arduino platform.

*Data collection from Arduino and BBC micro:bit weather stations via computer*

For data collection using a computer, we can advantageously use one of the freely available programs that can record data sent by serial transmission from weather stations via a USB connector into a text file. The latest version of the CoolTerm program, available at: https://coolterm.en.lo4d.com/windows, has proven to be the best for us. After downloading the program, just run it; there is no need to install it.

*2.2 VinciLab weather station*

The pressure, temperature and humidity sensors were installed in a 0.5 L plastic bottle and placed on a windowsill out of direct sunlight; the pressure sensor was left in the room. The measurement was carried out using a VinciLab data logger, set for long-term measurement (72 hours), with data recording every minute. After the measurement, the data was transferred to a computer and processed in the MS Excel environment. The arrangement of the weather station, as well as the placement of the sensors in the plastic bottle, is shown in Figure 5a.

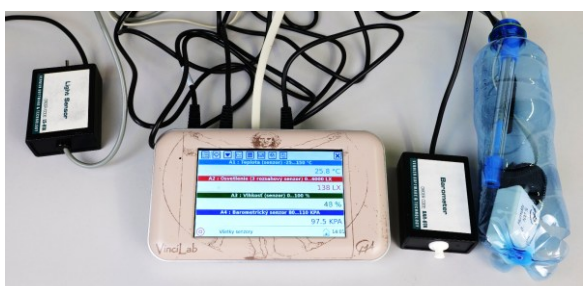
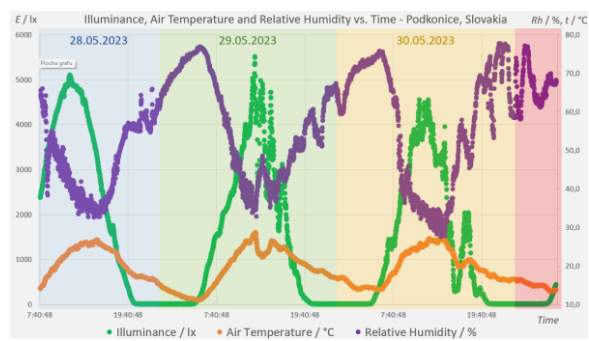

a)  b)

**Figure 5a, b.** Layout of the VinciLab weather station (a), Graph of illuminance, temperature of air and relative humidity vs. time (b).

## 3. Ideas for processing data from weather stations

After completing data collection, students can use the MS Excel spreadsheet to create graphs of the course of individual physical quantities over time (temperature, pressure, air humidity, light intensity) or their combinations. They can visually identify different changes or correlations between variables - e.g. a sudden change in pressure during the passage of an atmospheric front, an increase or decrease in humidity, fluctuations in light during the day, as well as correlations between different variables, e.g. a change in humidity when the temperature drops or rises, the correlation of light intensity is related to temperature, etc.

An example of how measured data can look in graphic form is shown in Figure 5b. The graph shows the values of light intensity, relative air humidity and air temperature measured over several days (the pressure changed very little during these days; therefore, its values are not given). Individual days are highlighted in colour to make it easier to navigate the graph. We can approximately determine the values directly from the graph, but more complete information can be obtained from the table of measured data.

As part of the discussion following the experiment, it is appropriate to create space for final reflection and ask questions that prompt students to think more deeply. Students can suggest ways to improve the weather station's technical capabilities, such as adding additional sensors (rain gauge, anemometer, $CO_2$ sensor), expanding memory options with an SD card, or ensuring autonomous operation without the need for a computer. They can also consider the practical applications of the weather station in the school environment, such as using it as a permanent device in the schoolyard, in a greenhouse, for monitoring the microclimate in classrooms, or as part of a broader environmental or geographical project. Finally, it is also possible to discuss ways of easier visualization and sharing of measured data, for example, through a web interface, online graphs, or exporting data to cloud storage, which would add new didactic and practical possibilities to the weather station.

## 4. Conclusion

Qualitative verification of the usability of different versions of the digital weather station in real classroom settings is planned for the end of 2025 within the DiTEdu national project. Therefore, students' guides and detailed methodological materials for teachers were prepared for each version of the weather station, containing detailed information on their construction and use.

The use of digital weather stations in teaching depends not only on practical factors such as students' age, available time, technical equipment of the school, and teachers' experience, but also on educational goals. If the aim is to strengthen cross-curricular links within STEM (e.g., connecting physics with computer science and environmental education), students can build and program the weather station themselves, which requires more time. If, however, the focus is primarily on working with measured data, it is more suitable to use ready-made stations that provide prepared datasets for analysis and interpretation.

Our preliminary experience, particularly from informal physics education organised at the authors' workplace, indicates that such weather stations have strong potential for developing students' skills in data collection, working with graphs and interpreting real-world measurements. Since both data collection and graph interpretation are key elements of scientific literacy, these activities can significantly contribute to students' understanding of physical phenomena. Based on our work with teacher training and student projects, we also expect that the construction of microcontroller-based weather stations will not only be attractive for high school students, but also for students of computer science and electrical engineering at secondary schools, especially in the context of longer-term projects.

Ultimately, a simple digital weather station proves to be an effective tool for fostering scientific literacy, critical thinking, and students' interest in technical sciences. Its design and programming promote logical and algorithmic thinking while also encouraging hands-on experimental work. To support teachers in adopting such innovative practices, we plan to continue providing resources and assistance through the electronic platform of the Physics Teachers' Club website: https://www.fpv.umb.sk/katedry/katedra-fyziky/klub-ucitelov-fyziky/.

## Acknowledgement

The presented methodology was developed as part of the national project *Digital Transformation of Education and School (DiTEdu)*, co-financed by the European Union. Martin Plesch acknowledges additional support from the project no. 09I03-03-V04-00425 of the Research and Innovation Authority (Next Generation EU).